\documentclass[10pt]{article}

\usepackage{amsmath,amsfonts,amssymb,amsthm,verbatim,mathrsfs,eufrak,pxfonts,txfonts,algorithmic,algorithm}
\usepackage[utf8]{inputenc}
\usepackage{graphicx} 
\usepackage{amsthm}
\usepackage{multirow}
\usepackage{geometry}
\usepackage{booktabs}
\usepackage{array}
\usepackage{paralist}
\usepackage{verbatim}
\usepackage{subfig}
\usepackage{amsmath, amsthm, amssymb}
\usepackage{authblk}

\usepackage{fancyhdr} 
\usepackage{sectsty}
\allsectionsfont{\sffamily\mdseries\upshape} 

\usepackage[nottoc,notlof,notlot]{tocbibind} 
\usepackage[titles,subfigure]{tocloft} 

\title{Inferring the Clonal Structure of Viral Populations from Time Series Sequencing}
\author[1]{Fotso-Chedom D}
\author[2]{Murcia PR}
\author[1,3]{Greenman CD\dag}
\affil[1]{The Genome Analysis Centre, Norwich Research Park, Norwich, UK}
\affil[2]{MRC-University of Glasgow Centre for Virus Research, UK}
\affil[3]{School of Computing Sciences, University of East Anglia, Norwich, UK}

\date{\vspace{-5ex}}

\begin{document}

\maketitle

\begin{abstract}
RNA virus populations will undergo processes of mutation and selection resulting in a mixed population of viral particles. High throughput sequencing of a viral population subsequently contains a mixed signal of the underlying clones. We would like to identify the underlying evolutionary structures. We utilize two sources of information to attempt this; within segment linkage information, and mutation prevalence. We demonstrate that clone haplotypes, their prevalence, and maximum parsimony reticulate evolutionary structures can be identified, although the solutions may not be unique, even for complete sets of information. This is applied to a chain of influenza infection, where we infer evolutionary structures, including reassortment, and demonstrate some of the difficulties of interpretation that arise from deep sequencing due to artifacts such as template switching during PCR amplification.
\end{abstract}

\let\thefootnote\relax\footnote{\dag Corresponding Author}


\section*{}

\vspace{-5ex}

RNA viruses have evolutionary dynamics characterized by high turnover rates, large population sizes and very high mutation rates \cite{beerenwinkel2011ultra}, \cite{duffy2008rates}, resulting in a genetically diverse mixed viral population \cite{beerenwinkel2011ultra}, \cite{skumsreconstruction}. Subpopulations in these mixtures containing specific sets of mutations are referred to as clones and their corresponding mutation sets as haplotypes. Unveiling the diversity, evolution and clonal composition of a viral population will be key to understanding factors such as infectiousness, virulence and drug resistance \cite{zagordi2012probabilistic}.

High throughput sequencing technologies have resulted in the generation of rapid, cost-effective, large sequencing datasets \cite{luciani2012next}. When applied to viruses, the set of reads obtained from a deep sequencing experiment represents a sample of the viral population which can be used to infer the underlying structure of that population at an unprecedented level of detail \cite{beerenwinkel2011ultra}.

In this study, we aim to identify the haplotypes of clones and quantify their prevalence within a viral population. The method also constructs evolutionary histories of the process consistent with the data. Reconstructing the structure of a mixed viral population from sequencing data is a challenging problem \cite{prosperi2011combinatorial}. Only a few works address the issue of viral mixed population haplotype reconstruction which infer both the genomes of sub-populations and their prevalence. Reviews of the methods and approaches dealing with these issues can be found in \cite{beerenwinkel2011ultra}, \cite{barzon2011applications}, \cite{pybus2009evolutionary}, \cite{beerenwinkel2012challenges} and \cite{zagordi2010deep}.

These works frequently make use of read graphs, which consist of a graph representation of pairs of mutations linked into haplotypes \cite{eriksson2008viral}. Haplotypes then correspond to paths through these graphs, although not every path will necessarily be realized as a genuine haplotype, which can lead to over-calling haplotypes. Different formalizations of this problem has led to different optimization problems in the literature \cite{beerenwinkel2012challenges}, including minimum-cost flows \cite{westbrooks2008hcv}, minimum sets of paths \cite{eriksson2008viral}, \cite{zagordi2011shorah}, probabilistic and statistical methods \cite{prosperi2011combinatorial}, network flow problems \cite{skumsreconstruction,astrovskaya2011inferring}, minimum path cover problems \cite{t2012haplotype}, maximizing bandwidth \cite{mancuso2011viral}, graph coloring problems \cite{huang2011qcolors} or K-mean clustering approaches \cite{eriksson2008viral}. After the haplotypes are constructed, in many cases an expectation-maximisation (EM) algorithm is used to estimate their prevalence in the sampled population. Some other works \cite{zagordi2012probabilistic}, \cite{prabhakaran2010hiv} use a probabilistic approach instead of a graph-based method.

In this work we take an integrative approach to address both the genetic diversity and the evolutionary trajectory of the viral population. The method presented is not read graph based and constructs evolutionary trees and recombination networks weighted by clone prevalences. This reduces the size of the solution set of haplotypes. The method does not rely exclusively on reads physically linking mutations so is applicable to longer segments. The method will also be shown to have particular utility with time series data and is highlighted on a chain of infections by influenza (H3N8).

The question of the influenza genome diversity has been addressed in the literature largely between strains or samples from different hosts, considering one single dominant genome for each host \cite{tsai2011influenza}. Within-host evolution is a source of genetic diversity the understanding of which may lead to the development of models that link different evolutionary scales \cite{pybus2009evolutionary}. Kuroda et al. \cite{kuroda2010characterization} addressed the question of evolution within a single host of influenza extracted on a patient who died of an A/H1N1/2009 infection, but with a focus on HA segment using a de-novo approach. Our approach provides a method to further understand within host evolution of such viruses.

The next section highlights the approach with an overview of the tree and network construction methods with simulated data, followed by an application of the method to a daily sequence of real influenza data. The Methods section describes the construction of the trees and networks in more detail.


\section*{Results}

We next outline the tree and network construction methods.

\subsection*{An Evolutionary Tree}

\begin{figure}[htbp]
\begin{center}
\includegraphics[width=13cm]{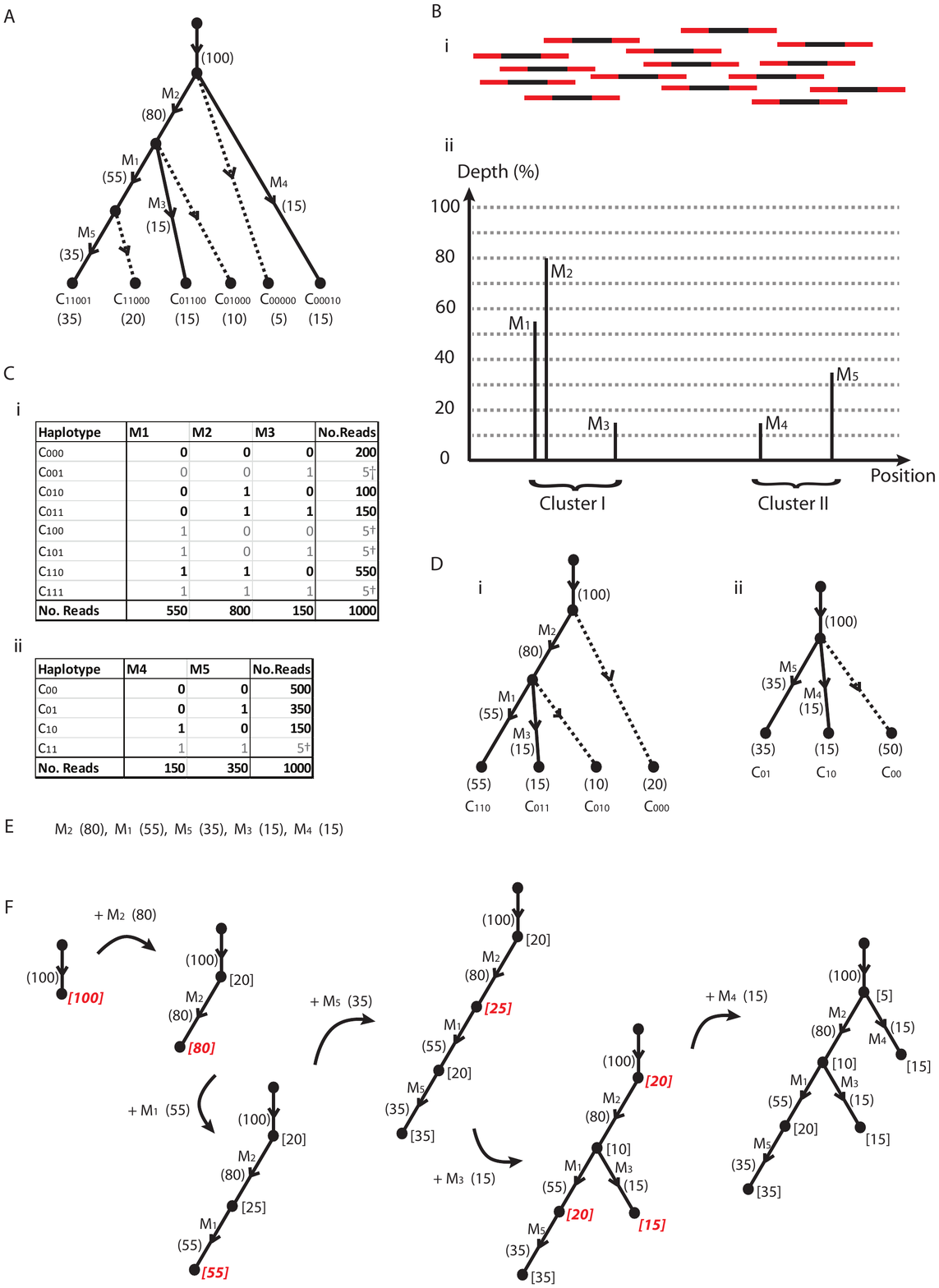}\caption{(A) A contrived evolution of a mixed viral population involving five mutations and six clones. Dotted lines indicate internal nodes extended to a leaf. (B) A notional representation of sequencing across the region of interest, and the resulting Depth-Position graph. Paired reads bridge two clusters of mutations. (C) Read count data obtained for the two clusters, with total depth x1000, along with artifacts \dag. (D) Evolutionary trees corresponding to (Ci,ii). (E) Ordered list of mutations and population prevalences (\%). (F) Reconstruction of original tree in (A).}
\label{PedagogicSample}
\end{center}
\end{figure}

Consider the pedagogic simulation in Figure \ref{PedagogicSample}, where we have a region of interest (such as an influenza segment, for example) that has undergone mutational and selective processes encapsulated by the evolution tree in Figure \ref{PedagogicSample}A. This tree contains five mutations $M_1,M_2,M_3,M_4,M_5$ that lie on various branches of the tree. These combine into the six clones that are the leaves of the tree. For example, the second leaf is labeled $C_{11000}$, indicating a clone with haplotype consisting of mutations $M_1,M_2$ but not $M_3,M_4,M_5$. Note that the path from the root of the tree to this leaf crosses the two branches corresponding to mutations $M_1$ and $M_2$. The number $20$ at the leaf indicates that this clone makes up $20\%$ of the viral population, and is termed the \emph{prevalence}.

Note that these prevalences form a \emph{conserved flow network} through the tree \cite{Papa}. For example, the prevalence of mutation $M_1$ is $55\%$, which accounts for the two haplotypes $C_{11001}$ and $C_{11000}$, with prevalences $35\%$ and $20\%$, respectively. In general, we find that the prevalence flowing into a node of the tree must equal the sum of the exiting prevalences. This represents conservation of the viral sub-populations. The total prevalence across all the leaves is therefore $100\%$.

In reality we are not privy to this information and perform a sequencing experiment to investigate the structure. This takes the form of molecular sequencing, where we detect the five mutations, which each have a different \emph{depth} of sequencing, as portrayed in Figure \ref{PedagogicSample}B. We will later see with real influenza data in that percentage depth can be reasonably interpreted as prevalence. Furthermore, we can look at the mutations arising on individual sequencing reads and group them into clusters. For our example, this groups the mutations into two clusters, giving the haplotype tables in Figure \ref{PedagogicSample}C.

We first construct an evolution tree for each of these tables. Our approach is based upon two sources of information; one utilizes mutation sequencing depth with a pigeon hole principle, the other utilizes linkage information from haplotype tables.

Now we have mutation $M_2$ present in $80\%$ of viruses and mutation $M_1$ present in $55\%$ of viruses. If these mutations are not both simultaneously present in a sub population of viruses, then the mutations are exclusive. This implies the two populations of size $80\%$ and $55\%$ do not overlap. However, the total population of viruses containing either of these viruses would then be greater than $100\%<80\%+55\%$. This is not possible, and the only explanation is that a subpopulation of viruses contain both mutations; the pigeon hole principle. The only tree-like evolutionary structure possible is that $M_1$ is a descendant of $M_2$, as indicated by the rooted, directed tree in Figure \ref{PedagogicSample}Di. Note that we have not utilized any haplotype information to infer this, just the mutation prevalence of the two mutations and a pigeon hole principle.

Mutation $M_3$ has a prevalence that is too low to repeat a prevalence based argument. However, we have a second source of information; the paired read data that can link together mutations into the haplotypes in Figure \ref{PedagogicSample}Ci. This table is based on three mutations, which group into $2^3=8$ possible haplotypes. However, a tree structure with three mutations will only contain four leaves \cite{semple2003phylogenetics} and we see that four of the halpotypes (emboldened) have notably larger counts of reads and are likely to be genuine. The four haplotypes with a notably lower read counts are likely to be the result of sequencing error at the mutant base positions, or template switching from a cycle of rtPCR, and are ignored. The presence of genuine haplotypes $C_{011}$ and $C_{010}$, lead us to conclude that $M_3$ is descendant from $M_2$ but not $M_1$, resulting in the tree of Figure \ref{PedagogicSample}Di.

From the mutation prevalences $55\%$, $80\%$ and $15\%$ of $M_1$, $M_2$ and $M_3$, we can also use the conserved network flow to measure the haplotypes prevalence. For example, the leaf descending from $M_2 (80\%)$, but not $M_1 (55\%)$ or $M_3 (15\%)$ (clone $C_{010}$ of Figure \ref{PedagogicSample}Di) must represent the remaining $10\%=80\%-55\%-15\%$ of the population.

This provides us with two sources of information (sequencing depth and linkage information) we can utilize to reconstruct the clone haplotypes, prevalence, and evolution. However, not all mutations can be connected by sequencing reads. They may be either separated by a distance beyond the library insert size, or may lie on distinct (unlinked) segments. Our approach is then as follows. We first construct a tree for each cluster of linked mutations. This will be a subtree of the full evolutionary structure. We then construct a supertree from this set of subtrees.

Now both of the trees in Figure \ref{PedagogicSample}D must be subtrees of a full evolutionary tree for the collective mutation set so we need to construct a supertree of these two trees. We can do this recursively as follows. We take the mutations and place them in decreasing order according to their prevalence, as given in Figure \ref{PedagogicSample}E. We then attach branches corresponding to these mutations to the supertree in turn, checking firstly network flow conservation, and secondly that the haplotype information in the subtrees is preserved. The steps for this example can be seen in Figure \ref{PedagogicSample}F. We start with a single incoming edge with prevalence $100\%$; the entire viral population. We next place an edge corresponding to $M_2$, the mutation with maximum prevalence of $80\%$. The next mutation in the tree either descends from the root or this new node. Any descendants of $M_2$ must have a prevalence less than this $80\%$. Any other branches must descend from the top node but can only account for up to $20\%$ of the remaining population. These two values are the \emph{capacities} indicated in square brackets. The next value we place is $M_1$ with prevalence $55\%$. This is beyond the capacity $20\%$ of the top node, so $M_1$ is descendant to $M_2$, accounting for $55\%$ of the $80\%$, leaving $25\%$. We thus have a three node tree with capacities $20\%$, $25\%$ and $55\%$. The third ordered mutation $M_5$ has prevalence $35\%$, which can only be placed at the bottom node with maximum capacity. Out next mutation $M_3$ has a prevalence $15\%$ that is less than any of the four capacities available, and no useful information on the supertree structure is obtained. This branch is the first to use haplotype information. We know from the first subtree that the corresponding branch is a descendant of $M_2$ and not $M_1$. The only node we can use (in red) has capacity $25\%$ and we place the branch. For the final branch corresponding to mutation $M_4$, the prevalence $15\%$ is less than four available capacities. The second subtree tells us that $M_4$ is not a descendant of $M_5$. This only rules out one of the four choices, and any of the three (red) nodes will result in a tree consistent with the data. The top node selected results in a tree equivalent to that in Figure \ref{PedagogicSample}A. To see this tree equivalence, the internal nodes in the last tree of Figure \ref{PedagogicSample}F have additional leaves attached (dotted lines) to obtain Figure \ref{PedagogicSample}A.

We thus find that a single dataset can result in several trees that are consistent with the data. However, having a time series of samples means a tree consistent with all days of data is required, which will substantially reduces the solution space. Note that the prevalences of the clones at the leaves of the tree results from this recursive process. We thus find that supertree construction is relatively straightforward with the aid of prevalence.

However, trees do not always fit the data. This can be due to recombination occurring within segments, or re-assortment occurring between segments. In the next section we construct recombination networks to cater for this, although we will see that they cannot be constructed as efficiently as trees.


\subsection*{A Recombination Network}

\begin{figure}[t!]
\centering
\includegraphics[width=12.2cm]{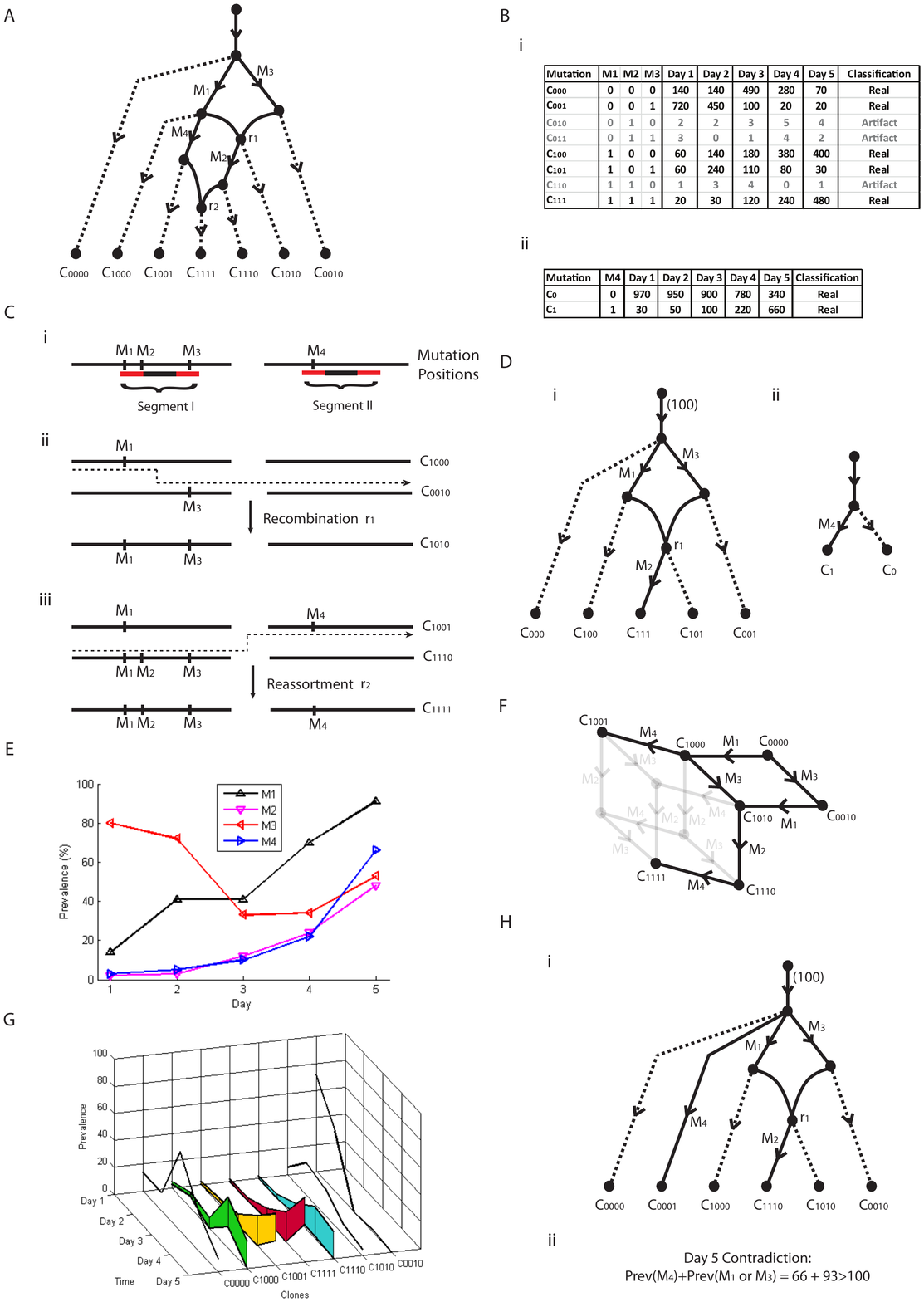}
\caption{(A) A pedagogic evolution recombination network with four mutations and two recombination events across two segments. (B) Typical haplotype tables arising from (A). (Ci) Two clusters of mutations grouped by paired reads on two segments. (Cii) Clones $C_{1000}$ and $C_{0010}$ undergo within segment recombination into $C_{1010}$, with a crossover site between $M_1$ and $M_3$. (Ciii) Clones $C_{1001}$ and $C_{1110}$ undergo between segment recombination (reassortment) into $C_{1111}$. (D) Recombination networks arising from the haplotype tables in (B). (E) The prevalence of the four mutations across five days. (F) Phylogenetic network associated with (A). (G) Point and range prevalence estimates. (Hi) A network consistent with the two networks of (D). (Hii) Incompatible prevalence conditions associated with (Hi).}
\label{NetworkFigure}
\end{figure}

In Figure \ref{NetworkFigure}A we see another simulated evolution based upon the two segments in Figure \ref{NetworkFigure}Ci that accumulate four mutations, $M1$, $M2$, $M3$ and $M4$. First we have mutations $M1$ and $M3$. Then we have the first of two recombination events, $r1$, where we have recombination within the first segment as described in Figure \ref{NetworkFigure}Cii. We then have mutations $M2$ and $M4$, followed by the second recombination event $r2$ in Figure \ref{NetworkFigure}Ciii, a re-assortment between the two segments. This results in the seven clones given at the leaves of Figure \ref{NetworkFigure}A. The  prevalences of the four mutations across five time points are given in Figure \ref{NetworkFigure}E. Note that we no longer have the conservation of prevalence observed in trees. For example, mutations $M1$ and $M3$ are on distinct branches extending from the root, yet their total prevalence is in excess of $100\%$ (on Day 5 for example). This is due to recombination $r1$ resulting in the presence of a clone containing both mutations. The use of the prevalence to reconstruct this structure from observable data thus requires more care. 

Now we see in Figure \ref{NetworkFigure}Ci that the four mutations cluster into two groups of mutations each bridged by a set of paired reads, resulting in two tables of read counts in Figure \ref{NetworkFigure}Bi,ii. We would like to reconstruct the evolution in Figure \ref{NetworkFigure}A from these data.

Firstly, we need to decide which of the haplotypes in Figure \ref{NetworkFigure}B are real. The haplotypes with consistently low entries are classified as artifact (in opaque). We next use a standard approach (such as a canonical splits network \cite{Huson}) to construct sub-networks from the real haplotypes in each of these tables, such as those given in Figure \ref{NetworkFigure}Dii. We then build super-networks ensuring that all sub-networks are contained as a sub-graph. There does not appear to be an efficient way of doing this (such as ordering by prevalence which works so well with trees) so a brute force approach is taken, where we construct all possible networks that contain four mutations and the haplotypes observed in Figure \ref{NetworkFigure}Bi,ii.

This results in many candidate super-networks. We now find that the prevalence information can be used to reject many cases. For example, the super-network in Figure \ref{NetworkFigure}Hi contains both sub-networks of Figure \ref{NetworkFigure}Di,ii as subgraphs. Note that the root node, representing the entire $100\%$ of the population, has daughter branches containing mutations $M4$, $M1$ and $M3$. However, from the prevalences on Day 5 we see that $M4$ has prevalence $66\%$ and $M1$ and $M3$ (which recombine) have a collective prevalence (from clones $C_{001}$, $C_{100}$, $C_{101}$,  and $C_{111}$ in Figure \ref{NetworkFigure}Bi) of $93\%$. This is in excess of the possible $100\%$ available and the network is rejected.

Application of filtering by prevalence (see Methods section for full details) rejects all networks with one recombination event, so we try all networks with two recombination events, resulting in just seven possible recombination networks. These all contain the same set of clones, all of which correspond to the single phylogenetic network in Figure \ref{NetworkFigure}F. Although only one recombination event is present across the subnetworks, all super-networks with one recombination event were filtered out and two recombination events were required.

Lastly we require estimates of the prevalences of each of the seven clones. We would like to match these to the prevalences in the tables of Figure \ref{NetworkFigure}B. This is a linear programming problem, the full details of which are given in the Methods section. The resulting estimates are given in Figure \ref{NetworkFigure}G where we see that some clones have point estimates, whereas others have ranges. For example, we see that clone $C_{0010}$ has a point estimate for each day. This is because it is the only clone of the super-network that corresponds to clone $C_{001}$ of Figure \ref{NetworkFigure}Bi and their prevalences can be matched. Conversely, we see ranges for the prevalences of clones $C_{1110}$ and $C_{1111}$. This is because both clones correspond to clone $C_{111}$ of Figure \ref{NetworkFigure}Bi and prevalence estimates for each clone cannot be uniquely specified.

Full details of this approach can be found in the Methods section. In the next section we describe the results obtained when applying these methods to a time series of influenza samples.


\subsection*{Application to Influenza}
\label{FluSection}

The data used in this study were generated from a chain of horse infections with influenza A H3N8 virus (Murcia, unpublished). An inoculum was used to infect two horses labeled 2761 and 6652. These two animals then infected horses labeled 6292 and 9476. This latter pair then infected 1420 and 6273. The chain continued and daily samples were collected from the horses resulting in 50 samples in total. For the present study we used 16 samples; the inoculum and hosts 2761 (days 2 to 6), 6652 (days 2, 3 and 5), 6292 (days 3 to 6) and 1420 (days 3, 5 and 6).

\begin{figure}[htbp]
\begin{center}
\includegraphics[width=13cm]{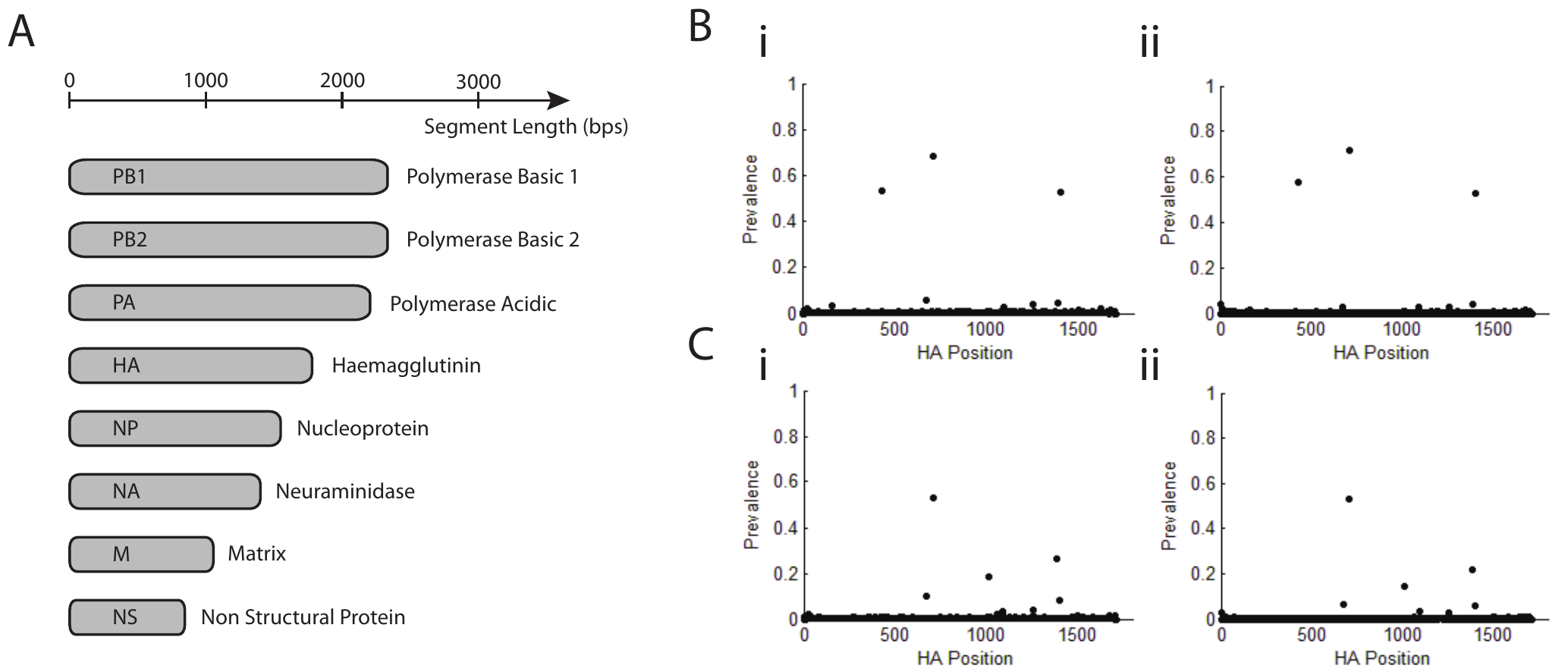}
\caption{(A) Size and function of the eight flu segments. (B,C) Depth of mutations from segment 4 from host 2761 Day 4 and host 6292 Day 3, respectively. (i,ii) Results from Hi-Seq and Mi-Seq experiments on separate libraries from the same samples.}
\label{FluSegments}
\end{center}
\end{figure}

Influenza  A virus is a member of the family Orthomyxoviridae which contains eight segmented, negative-stranded genomic RNAs commonly referred to as segments and numbered by their lengths from the longest 2341 to the shortest 890 bps  \cite{tsai2011influenza}, as summarized in Figure \ref{FluSegments}A.

Daily samples were collected from each host and paired end sequenced was performed with Hi-seq and Mi-seq machines.  The samples sequences were aligned with Bowtie2 \cite{langmead2012fast} with default parameters. We obtain for each sample a SAM file containing mapping information of all the different reads in the sample. Any mapped read whose average Phred-quality per base was less than $30$ were discarded.

In order to identify mutations from real data we need a reference sequence to compare the read sequences to. Consistent differences between the two can then be classified as a mutation. We constructed a majority consensus sequence from the inoculum sample. This consensus sequence was then used as a starting reference for the chain of infected animals. 

An amplification procedure was used to produce viral DNA. This involves PCR, which will result in different levels of amplification and mutations. This in turn is likely to introduce significant differences between the sequencing depth and prevalence. To combat this, all identical paired end reads (with equal beginning and endpoints) were grouped, classified as a single PCR product, deriving from a single molecule and only counted once. The depth of sequencing with these adjusted counts then provided an improved measurement of the prevalence of viral subpopulations. We compared an identical sample that was sequenced separately, the results of which can be seen across two samples in Figures\ref{FluSegments}B,C i,ii. Both the position and prevalence of mutations were reproducible to good accuracy suggesting proportional sequencing depth is a good surrogate for prevalence. 

We then applied the methods to sets of high prevalence mutations in each of the eight segments individually, and also to a set of three mutations from distinct segments. The main observations are below.


\subsection*{Within Segment Evolution}
\label{Recombination}

\begin{figure}[t!]
\begin{center}
\includegraphics[width=15cm]{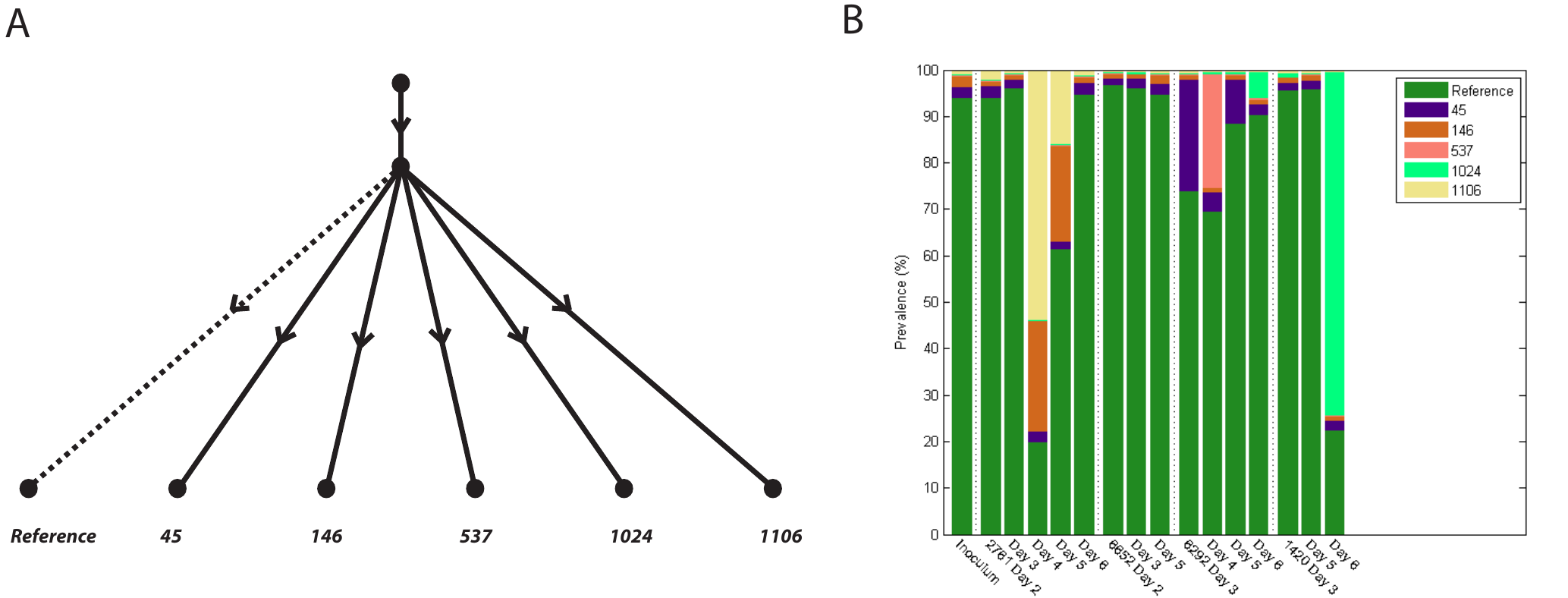}
\caption{(A) Evolution tree arising from five mutations on segment 6. (B) Prevalences of the clones across the times series.}
\label{MiSeqHiSeq}
\end{center}
\end{figure}

For segments 1, 3, 5, 6, 7 and 8 we obtained tree like evolutions for the segments. In all cases the mutations involved lay on distinct branches and were indicative of mutations arising in independent clones. Segment 6 can be seen in Figure \ref{MiSeqHiSeq}A, where we see five mutations on six branches. We also see from the stacked bar chart in Figure \ref{MiSeqHiSeq}B that many of the mutations arose during different periods in the infection chain.

However, the evolution structure of mutations within segments did not always appear to be tree like, with segment 2 containing one putative recombination event and seven mutations, and segment 4 containing three putative recombination events and six mutations. This latter case arose because we found three pairs of mutations in putative recombination. Using nucleotide positions as labels, these were (431, 674), (431, 709) and (709, 1401). That is, we found significant counts of all four combinations of mutations, labeled $C_{00}$, $C_{01}$, $C_{10}$ and $C_{11}$, lying on paired reads. Examples of typical counts for three (out of sixteen) samples are given in the top table in Figure \ref{TemplateSwitching}B (see Supplementary Information for full details). If the evolution is tree-like, reads from one of the types $C_{01}$, $C_{10}$ or $C_{11}$ should only arise as an artifact. Note that we have high read counts of all four categories, which is indicative of recombination.

However, various studies have shown that there is very little evidence of genuine recombination that occurs within segments of influenza \cite{Boni}, \cite{Boni2}, \cite{Boni3}, and these kind of observations can arise from template switching across different copies of segments during the rtPCR sequencing cycle \cite{TemplateSwitching}. We developed an analytic approach to consider this possibility in more detail.

Now if the true underlying structure is tree-like, it suggests that one of $C_{11}$, $C_{01}$ or $C_{10}$ arises purely from template switching (the wild type $C_{00}$ is assumed to always occur). This gives us the three models (labeled i-iii in Figure \ref{TemplateSwitching}A,B) to consider. We let $a$, $b$ and $c$ be the population proportions of the three real genotypes. We let $n$ be the probability that a cycle of rtPCR causes template switching. We then treat template switching as a continuous time three state random process. This allows us to derive probabilities that genotypes $C_{00}$, $C_{01}$, $C_{10}$ and $C_{11}$ arise on paired end reads, as given in Figure \ref{TemplateSwitching}A (see Methods for derivations). The counts of the four classes of read then follow a corresponding multinomial distribution. Maximum likelihood was used to estimate parameters, obtain log-likelihood scores, and a chi-squared measure of fit was obtained for each of the three models.

For the pair (431, 674) we found that the best log-likelihood, on all sixteen sampled days, was Model $1$ (Figure \ref{TemplateSwitching}i), where reads of type $C_{11}$ are artifacts arising from template switching alone. The parameters obtained provided an almost perfect fit; the expected counts were almost equal to the observed counts and the goodness of fit significance values were close to $1$. The other two models had substantially lower likelihoods and significantly bad fits. This tells us that if the underlying structure is a tree, it involves the three genotypes $C_{00}$, $C_{01}$ and $C_{10}$ and mutations 431 and 674 lie on distinct branches.

For the pair (431, 709) we found that the best log-likelihood, over the sixteen sampled days, was Model 3 (Figure \ref{TemplateSwitching}iii), where reads of type $C_{10}$ are artifacts. The parameters obtained provided an almost perfect fit on most days with goodness of fit significance values close to $1$. A couple of days had relatively poor fits, but were not significant when multiple testing across all sixteen days was considered. The model with $C_{11}$ as an artifact had very similar likelihoods, but the data exhibited significantly poor fits on multiple days. The model with $C_{01}$ as an artifact performed very badly. This tells us that if the underlying structure is a tree, it involves the three genotypes $C_{00}$, $C_{01}$ and $C_{11}$ and mutation 431 is a descendant of 709.

For the pair (709, 1401) we found that the best log-likelihood, on all sixteen sampled days, was the Model 2 (Figure \ref{TemplateSwitching}ii), where reads of type $C_{01}$ are artifacts. The parameters obtained provided an almost perfect fit on all days with goodness of fit significance values close to $1$. The other two models performed very badly. This tells us that if the underlying structure is a tree, it involves the three genotypes $C_{00}$, $C_{10}$ and $C_{11}$ and mutation 1401 is a descendant of 709.

\begin{figure}[t!]
\begin{center}
\includegraphics[width=15cm]{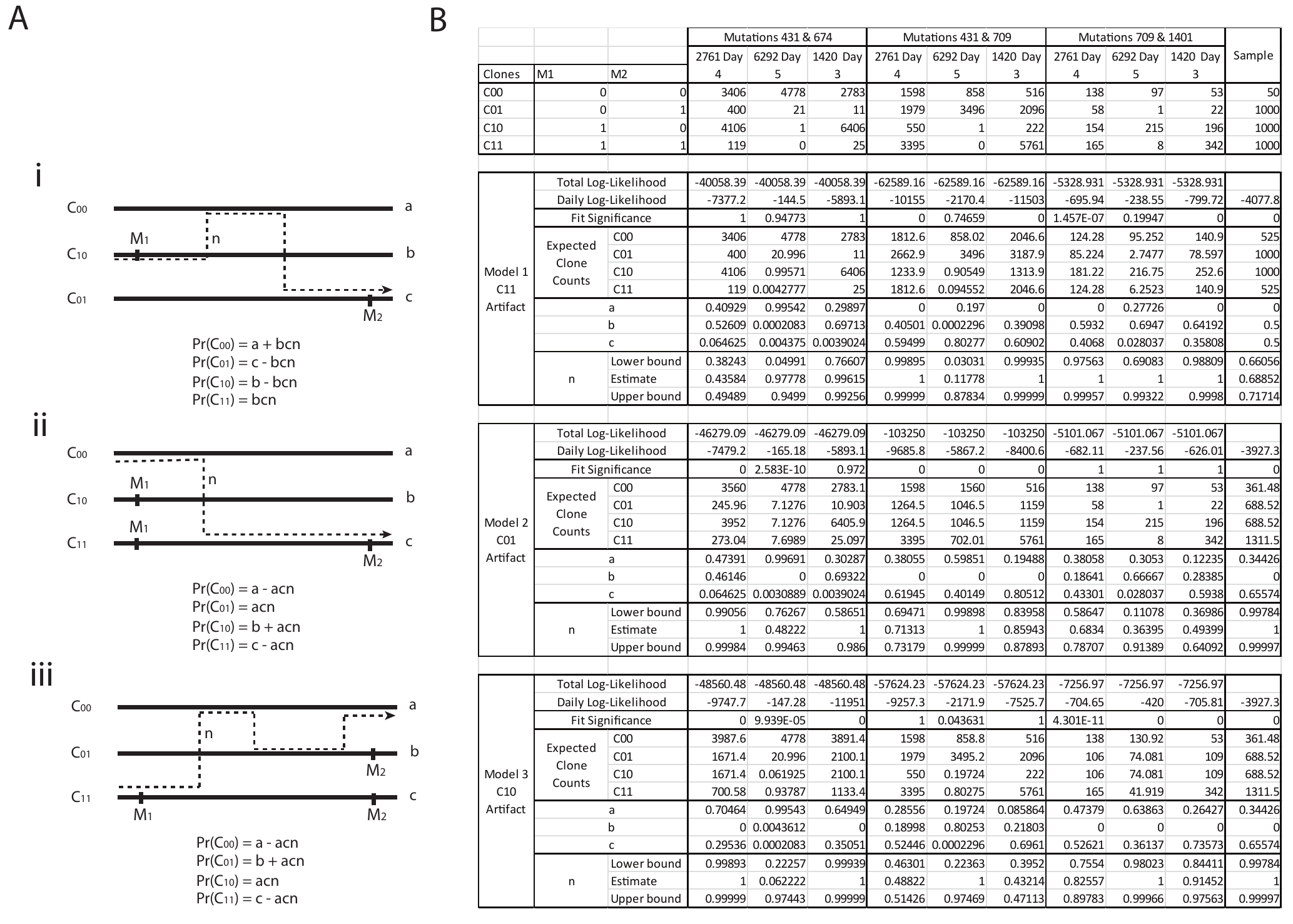}
\caption{(A) Three template switching models. (B) Fitted models for three pairs of linked mutations on segment 4.}
\label{TemplateSwitching}
\end{center}
\end{figure}

\begin{figure}[t!]
\begin{center}
\includegraphics[width=14.5cm]{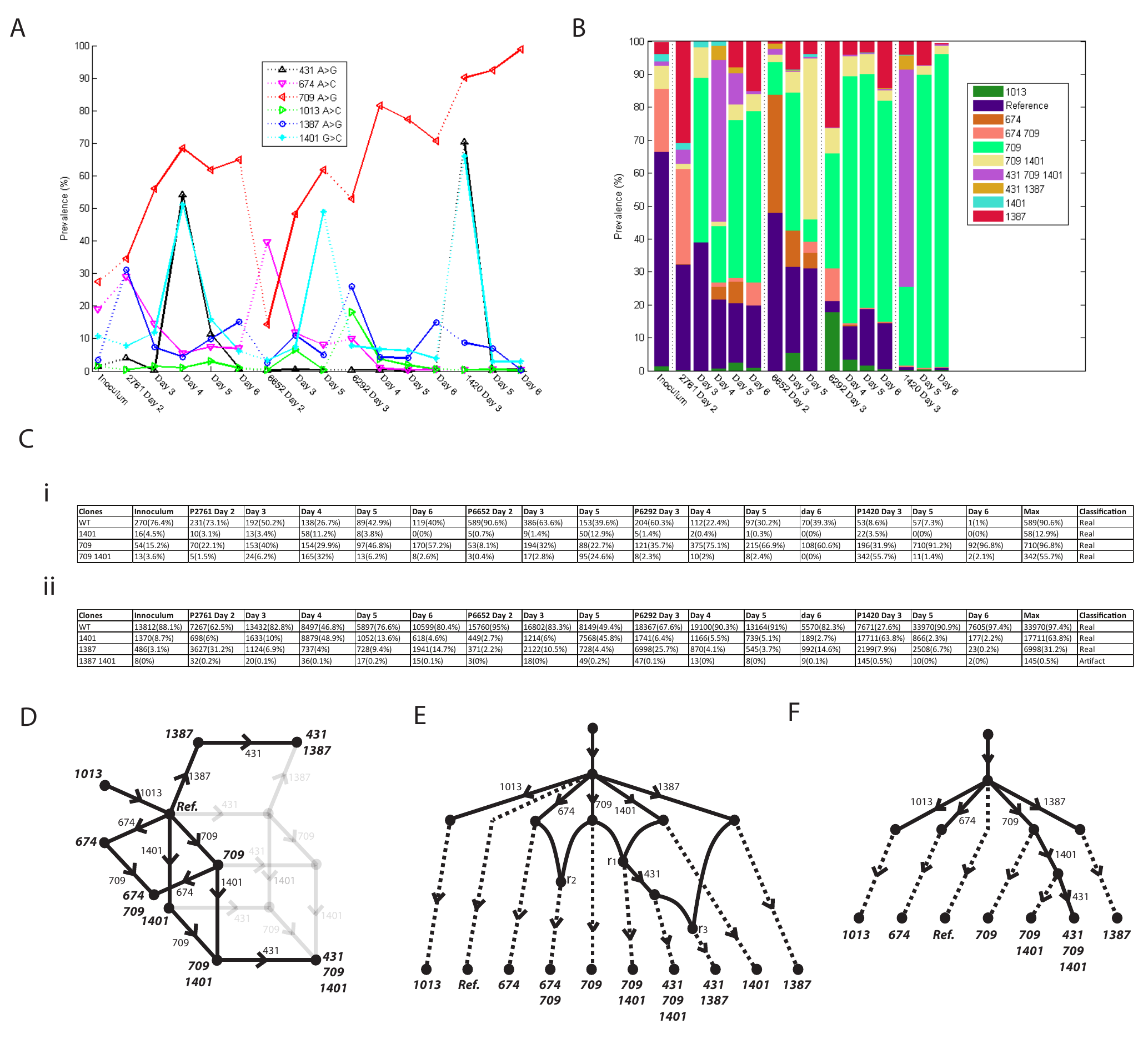}
\caption{(A) Mutation prevalences of six mutations in segment 4. (B) Prevalences of the ten associated clones. (C) Two tables of linked mutations exhibiting network like relationship of mutations 709 and 1401, and tree like relationship of 1387 and 1401. (D) The phylogenetic network of the single fitted cloneset. (E) One of 22 possible recombination networks that arise from (D). (F) Probable tree structure from (E) after template switching is considered.}
\label{MiSeqHiSeq2}
\end{center}
\end{figure}

Thus the three cases where data are indicative of recombination can be explained purely by template switching during rtPCR. This is reinforced somewhat by the fact that the same model emerged across all sampled days for each mutation pair. However, this does not definitively rule out recombination, which could also exhibit these consistent patterns across sampled days, and so care is needed when interpreting data. Furthermore, the rates of template switching required to explain the data without recombination were not always consistent. For example, in the sample from host 2761 Day 4, the estimated template switching between mutations (431,674) was $43.6\%$ ($95\%$ c.i $38.2\%$ - $49.2\%$). Between mutations (431,709) it was $48.8\%$ ($95\%$ c.i $46.3\%$ - $51.4\%$), giving reasonable agreement. Between mutations (709, 1401) it was somewhat higher, at $58.6\%$ ($95\%$ c.i $68.3\%$ - $78.7\%$), although this may be expected due to the greater distance between the mutations. However, in sample 1420 Day 3, the template switching rate for the pair (431,674), at $99.6\%$ ($95\%$ c.i $76.6\%$ - $99.3\%$), was notably higher than both the mutation pair (431,709), at $43.2\%$ ($95\%$ c.i $39.5\%$ - $47.1\%$), and mutation pair (709,1401), at $49.3\%$ ($95\%$ c.i $36.9\%$ - $64.1\%$). Although differences between samples (and so library preparations) may be expected, differences such as this in the same library are harder to explain without implicating genuine recombination.

We thus have two explanations of the data; genuine recombination or template switching artifacts. We consider both cases and then draw comparisons.

Firstly we consider segment 4 assuming recombination has taken place. The results can be seen in Figure \ref{MiSeqHiSeq2}. The prevalences of six mutations of interest are given in Figure \ref{MiSeqHiSeq2}A. Reasonable linkage information was available across the segment, including the two haplotype tables in Figure \ref{MiSeqHiSeq2}C. The first is linkage information between mutations 709 and 1401, where all four combinations of mutation occur to reasonable depth, implying recombination between the mutations. The second is between mutations 1387 and 1401, where we see only three haplotypes occur to significant depth, suggesting a tree like evolutionary structure between the two mutations. The full set of tables is in Supplementary Information. Although the sequencing depth in the first table is lower, due to the rarer occurrence of sufficiently large insert sizes, the information gleaned is just as crucial. The most parsimonious evolution found involved three recombination events, resulting in the single cloneset contained in the phylogenetic network given in Figure \ref{MiSeqHiSeq2}D. There were $22$ possible recombination networks that fit this phylogenetic network, one example of which is given in Figure \ref{MiSeqHiSeq2}E. The relatively complete linkage information resulted in point estimates for the clone prevalences (rather than ranges), as given in Figure \ref{MiSeqHiSeq2}B. 

If we now assume that the recombination like events are template switching during rtPCR, then from above, we observed that mutations 431 and 674 are on distinct branches, mutation 431 is a descendant of 709, and 1401 is also a descendant of 709. This resolves all three reticulation events in the network of Figure \ref{MiSeqHiSeq2}E and we end up with the tree given in Figure \ref{MiSeqHiSeq2}F. However, this structure still has two minor conflicts. Firstly, the tree like structure suggests that mutation 431 should have a lower prevalence than 1401, and on most days it does. However, the sample from host 2761 Day 4 has prevalences $54.3\%$ and $51.2\%$ for mutations 431 and 1401, respectively. Similarly, the samples from host 1420 Day 3 are $66.1\%$ and $70.3\%$, respectively. Secondly, the four mutations 674, 709, 1013 and 1401 all descend from the root on separate branches and should have a total prevalence that is less than $100\%$ and on fifteen of sixteen samples this is true. However, on sample 6292 Day 3 the prevalences are $9.9\%$, $52.8\%$, $18.0\%$ and $26.0\%$, which combine to $106.7\%$. Although the conflicts are relatively small, these differences are larger than would be expected from Poisson sampling of such deep data. However, this is the most plausible tree structure we found.


\subsection*{Detecting Reassortment}
\label{Reassortment}

Re-assortments occur when progeny segments from distinct viral parents are partnered into the same viral particle, resulting in a recombinant evolutionary network. 

Now re-assortment is a form of recombination. This is usually possible to detect in diploid species such as human because linkage information is available across a region of interest, such as a chromosome, and recombination can be inferred. Furthermore human samples have distinct sequencing samples for each member of the species. Inferring re-assortment across distinct viral samples is more difficult because firstly we do not have linkage information across distinct segments, and secondly, we have mixed populations within each sample.

\begin{figure}[t!]
\begin{center}
\includegraphics[width=12.8cm]{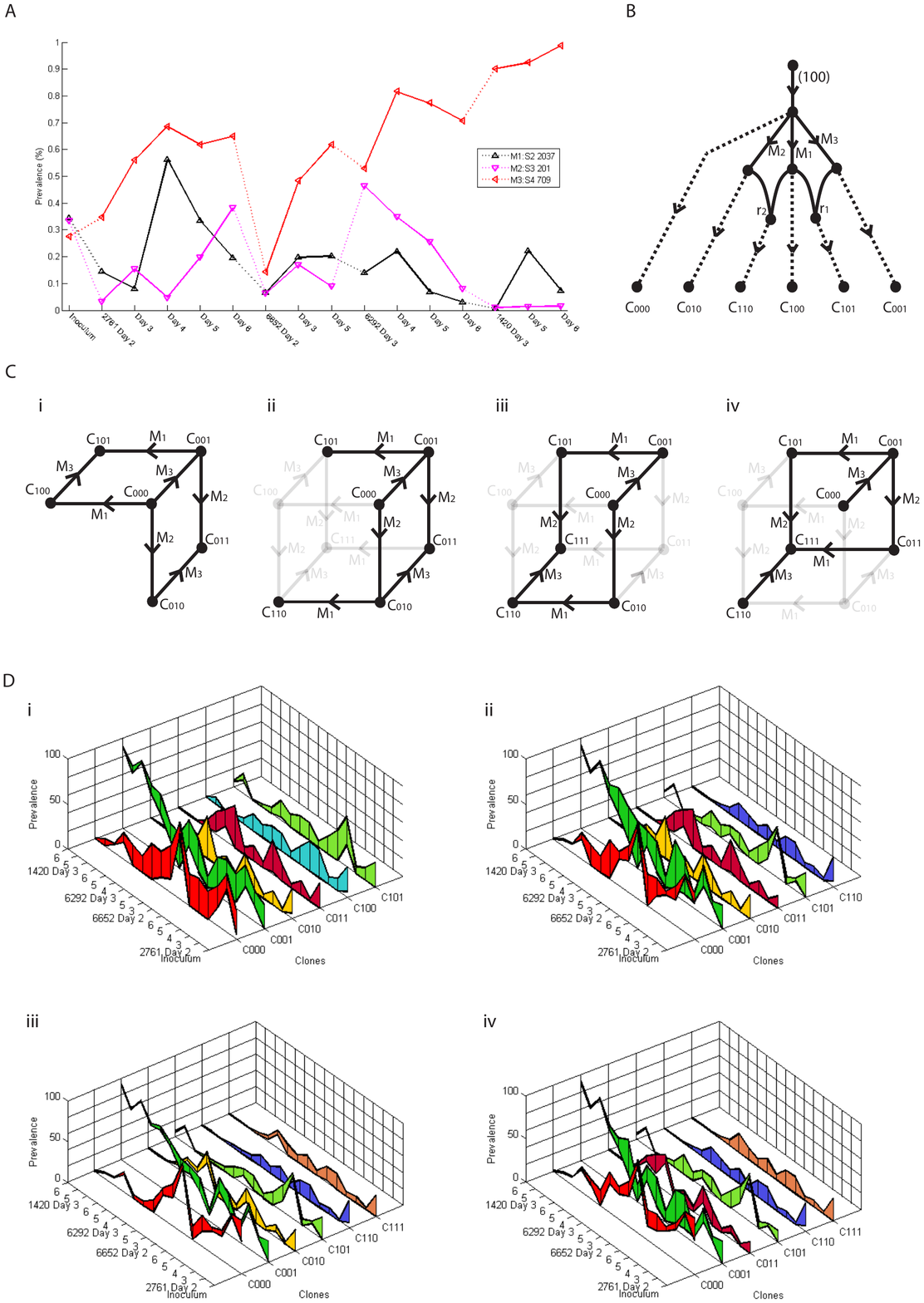}
\caption{Three mutations between three segments that indicate two re-assortment events. (A) Mutation prevalences across time series. (B) One of 51 recombination networks that fit the data. (C) Two phylogenetic networks that fit the data ((i) and (ii)-(iv)), corresponding to four clonesets. (D) Prevalence ranges for the four clonesets.}
\label{Reassortment}
\end{center}
\end{figure}

However, we show that re-assortment can still be detected within mixed population viral samples with the aid of information provided by prevalence. Consider Figure \ref{Reassortment}. We have three mutations in segments $2$, $3$ and $4$, along with their mutation nucleotide positions 2037, 201 and 709, respectively. We refer to the mutations as $S2\_2037$, $S3\_201$ and $S4\_709$ accordingly. We see in Figure \ref{Reassortment}D that $S2\_2037$ and $S3\_201$ have prevalences that alternate in magnitude across the 16 days sampled. If we assume a tree like structure, these two mutations cannot lie on a single branch, because one prevalence would have to be consistently lower than the other; they must therefore lie on distinct branches. Now mutation $S4\_709$ can; i) be on a distinct third branch, ii) be a descendant of $S2\_2037$, iii) be a descendant of $S3\_201$, iv) be an ancestor of $S2\_2037$, v) be an ancestor of $S3\_201$, or vi) be an ancestor of both. We can rule out all of these choices as follows.

Firstly we note that $S4\_709$ has a prevalence that is consistently larger than that of $S2\_2037$ or $S3\_201$, so cannot be a descendant of either mutation, ruling out ii) and iii). We see from sample 6292 Day 3 that $S2\_2037$ and $S3\_201$ have a total prevalence greater than $S4\_709$, meaning $S4\_709$ cannot be an ancestor of both mutations, ruling out vi). In this sample, the total prevalence of all three mutations is in excess of $100\%$, ruling out i). Now if $S4\_709$ and $S2\_2037$ lie on distinct branches, we see from 2761 Day 4 that their combined prevalence is in excess of $100\%$, ruling out v). Finally, if $S4\_709$ and $S3\_201$ lie on distinct branches, we see from 6292 Day 4 that their combined prevalence is in excess of $100\%$, ruling out iv). No tree structure is possible and we conclude the presence of re-assortment as the most likely explanation.

In fact, application of the full method reveals that two re-assortment events are required to explain the data. This results in 51 possible recombination networks, one such example is given in Figure \ref{Reassortment}B. These correspond to the four clonesets given in Figure \ref{Reassortment}C, arising from two possible phylogentic networks. The four clonesets have prevalences that could not be uniquely resolved; their possible ranges are shown in Figure \ref{Reassortment}D. Although we cannot uniquely identify the network or the prevalences, all solutions involved two re-assortments, one involving mutations $S4\_709$ and $S2\_2037$, the other involving $S4\_709$ and $S3\_201$. This observation was only possible because of inferences made with the prevalence.


\section*{Discussion}

We have introduced a methodology to analyze time series viral sequencing data. This has three aims; to identify the presence of clones in mixed viral populations, to quantify the relative population sizes of the clones, and to describe underlying evolutionary structures, including reticulated evolution.

We have demonstrated the applicability of these methods with paired end sequencing from a chain of infections of the H3N8 influenza virus. Although we could identify underlying evolutionary structures, some properties of the viruses and the resulting data make interpretation difficult. In particular, template switching during the rtPCR cycle of sequencing an RNA virus is known to occur, and can result in paired reads that imply the presence of recombination. Although any underlying tree like evolutions can still be detected, these artifacts confound the signal of any genuine recombination that may be taking place, making it harder to identify. The prevalence of mutations, measured as sequencing depth proportion, offers an alternative source of information that can help resolve these conflicts in theory, although more work is needed to evaluate how robust this metric is in practice.

For example, although tree like evolutions were identified in six of the segments, in the two remaining segments the approach found reticulated networks, with three distinct reticulated nodes in the hemagglutinin segments network. Although each of these nodes were consistent with template switching artifacts, the resultant tree structure could not quite be fitted to the mutation prevalences. Although this conflict implies the original network is correct and recombination has taken place, within segment recombination in influenza is rare \cite{Boni}, \cite{Boni2}, \cite{Boni3} and other explanations may be required. In particular, we note in Figure \ref{FluSegments}B that there are slight differences between the prevalences obtained from independent Mi-seq and Hi-seq runs. Although some of this will be due to Poisson variation of depth, there could be some biases in PCR over certain mutations, for example. The application of prevalence thus needs to be used with caution, and further studies are needed to fine tune this type of approach.

When the approach was applied to mutations in distinct segments, two re-assortment events were inferred. The differences in mutation prevalences were more marked in this case suggesting the inference is more robust and re-assortment more likely to have taken place. This is also biologically more plausible, with events such as this accounting for the emergence of new strains. We note that although re-assortment may have genuinely taken place, only one of the original clones (containing just mutation 709 on segment 4) survived the infection chain and a longitudinal study would not have picked up such transient clonal activity.

These methods utilized paired end sequencing data and showed that even when paired reads do not extend the full length of segments, or bridge distinct segments, we can still make useful inferences on the underlying evolutionary structures. The two main sources of information are the linkage offered by two or more mutations lying on the same paired reads, and the prevalence information. It is by utilizing the variability of the prevalence in a time series dataset that we can narrow down the predictions to a useful degree; application of this method to individuals days will likely result in too many predictions to be useful. Furthermore, this has greatest application to mutations of higher prevalence; this places more restrictions on possible evolutions consistent with the data. Subsequently, this kind of variability is most likely to manifest itself under conditions of differing selectional forces. A stable population is less likely to contain mutations moving to fixation under selective forces. Lower prevalence mutations will result, meaning less predictive power. Simulations also suggest that although clone-sets may be uniquely identified, prediction of the underlying reticulation network is difficult, with many networks explaining the same dataset.

As we lower the minimum prevalence of analyzed mutations, their number will increase. The number of networks will likely explode and raise significant challenges. Furthermore, single strand RNA viruses such as influenza mutate quickly, suggesting a preponderance of low prevalence mutations likely exist. This is further exacerbated by the fact that sequencing uses rt-PCR, introducing point mutations and template switching artifacts that create noise in the data. These processes are likely responsible for the grass-like distribution of low prevalence mutations visible in Figure \ref{FluSegments}B,C. Thus as we consider lower prevalence mutations we are likely to get a rapidly growing evolution structure of increasingly complex topology. The methods we have introduced, however, can provide useful information at the upper-portions of these structures.

The software ViralNet is available at www.uea.ac.uk/computing/software. The raw data is available from the NCBI (project accession number SRP044631). More detailed outputs from the algorithm are available in Supplementary Information.


\section*{Methods}
\label{Alg}

We now consider tree and network construction methods, a template switching model, and validation of the methods.


\subsection*{Tree Construction}
\label{AlgSection}

The construction of phylogenetic trees is a well established area \cite{semple2003phylogenetics}. Trees are frequently constructed from tables of haplotypes of different species. However, we have two properties that change the situation. Firstly, if we have a set of $n$ mutations linked by reads, we can have up to $2^n$ distinct haplotypes. However, a consistent set of splits from such a table should only have up to $n+1$ distinct haplotypes, in a split-compatible configuration \cite{semple2003phylogenetics}. To construct a phylogenetic tree we thus need to classify the genotypes as real or artifact. Secondly, we have prevalence information, in the form of a conserved network flow through the tree. This can help us to both decide which haplotypes to believe and to construct a corresponding tree. 

To describe the algorithm we first introduce some notation. Now, the evolutionary structure is represented by two types of rooted directed tree; one where each edge represents a mutation, such as in Figure \ref{PedagogicSample}F, and one where all leaves represent clones in the population, such as in Figures \ref{PedagogicSample}A,D. The first is a subtree of the latter. The latter has a conserved flow network. These will be termed the \emph{Compact Prevalence Tree} and \emph{Complete Prevalence Tree} respectively.

Now to each edge $e$ in the compact prevalence tree, we assign \emph{prevalence} $\rho(e)$. This represents the proportion of population containing the mutation represented by the edge $e$. The single directed edge $e_{in}(v)$ pointing toward a vertex $v$ (away from the root) represents a viral population of prevalence $\rho(e_{in}(v))$, all containing the mutation corresponding to edge $e_{in}(v)$, along with its predecessor mutations. The set of daughter edges $E_{out}(v)$ leading away from node $v$ represent populations containing subsequent mutations, each with prevalence $\rho(e), e \in E_{out}(v)$. The remaining population from $\rho(e_{in}(v))$ contains just the original mutation set, having a prevalence described by the \emph{capacity} $\zeta(v)$. The conservation of prevalence satisfied by each vertex $v \in T$ is then represented by the condition:

\begin{equation}
\rho(e_{in}(v)) = \zeta(v) + \sum_{e \in E_{out}(v)} \rho(e(v))
\label{PrevalenceConservation}
\end{equation}

The root node has total prevalence of $1$, representing the entire population of interest.

This describes the mutation based trees such as that in Figure \ref{PedagogicSample}F. To obtain a complete tree containing all the clones, we need to extend an edge from each internal node to represent the associated clone (these are the dashed lines in Figure \ref{PedagogicSample}A). The prevalence of the additional edges are equal to the capacities of the parental nodes.

We saw in Figure \ref{PedagogicSample}B that mutations can be clustered together, and evolution trees constructed for each cluster. We refer to these as \emph{Subtrees}. We then look for a tree that contains all such subtrees as a subset of edges. We refer to these as \emph{Supertrees}.

The algorithm is broken into two steps. The first calculates subtrees. The second calculates supertrees.

\begin{figure}[t!]
\centering
\includegraphics[width=11cm]{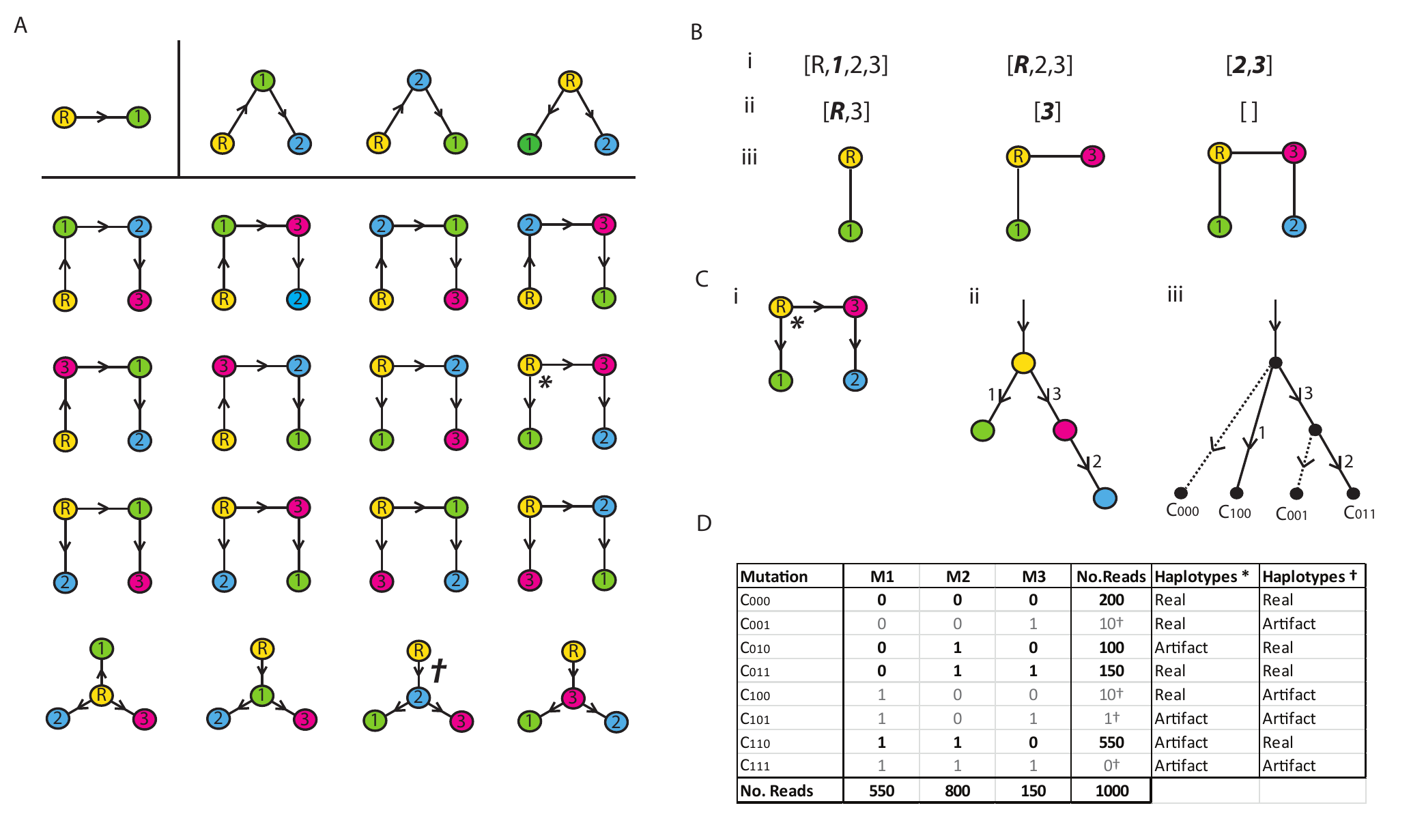}
\caption{(A) Cayley trees for $2$, $3$ and $4$ vertices. (Bi) Vertex list {\bf v} for example (*). (Bii) Pr\"{u}fer sequence {\bf p}. (Biii) Tree construction. (Ci) The graph directed away from the root. (ii) The equivalent compact clonal tree. (iii) The corresponding complete clonal tree. (D) Alignment of trees * and $\dag$ to haplotype tables.}
\label{Cayley}
\end{figure}

{\bf Step 1} \emph{Subtree Construction} Now, for $n$ mutations we have $2^n$ possible haplotypes, with corresponding counts $n_i, i=1,2,...,2^n$, and a tree with $n+1$ haplotypes to fit. This implies that $2^n-n-1$ of those counts are artifacts. For example, in Figure \ref{Cayley}D we see the simulated counts for $2^3$ haplotypes on $3$ mutations. Now Cayley's formula states that there are $n^{n-2}$ different labeled trees that can be constructed on n vertices \cite{Cayley}. These are easily constructed with the aid of Pr\"{u}fer sequences \cite{prufer}, which are any integer sequence $[p_1,...,p_{n-2}]$ such that $p_i \in \{1,2,...,n\}$. The first few examples are given in Figure \ref{Cayley}A.

To construct a tree we start with ${\bf p}=[p_1,...,p_{n-2}]$ and the vector ${\bf v}=[1,2,...,n]$. At each step we take the lowest entry of ${\bf v}$ not in ${\bf p}$, and the lowest entry of ${\bf p}$, and join the two corresponding nodes together with an edge. For example in Figure \ref{Cayley}B we start with ${\bf v}=[R,1,2,3]$, where the root node $R$ is treated as the minimum value, along with Pr\"{u}fer sequence ${\bf p}=[R,3]$. The smallest element of ${\bf v}$ not in ${\bf p}$ is $1$. The corresponding node is then joined to the node for the smallest element $R$ of ${\bf p}$, such as exemplified in Figure \ref{Cayley}Biii. These two elements are removed from ${\bf p}$ and ${\bf v}$ and the process repeated until we are left with two elements in ${\bf v}$. Our example leaves us with the two elements $2$ and $3$, the corresponding nodes of which are then joined by an edge. The edges are then directed away from the root, resulting in the prevalence clonal tree in Figure \ref{Cayley}Cii. The corresponding complete prevalence tree is in Figure \ref{Cayley}Ciii.

Once we have all the possible subtrees constructed, we use maximum likelihood to select the most plausible tree. Consider, for example, the penultimate column of Figure \ref{Cayley}D, which correspond to the four haplotypes for the tree $(*)$ in Figure \ref{Cayley}A-C. Note that the haplotype $C_{110}$ with a count 550 is an artifact for this tree. If each mutation artifact arises with probability $\epsilon$, then an artifact read of type $C_{110}$ contains two mutant bases and occurs with probability $\epsilon^2$. We can then construct log-likelihoods (summed across time points) for the artifact counts arising from clones that do not belong to the putative tree being tested. We then assume Poisson distributed counts and construct the following likelihood function for a given putative clonal tree $\mathscr{T}$:

\begin{equation}
\label{eq:Fscore}
\mathscr{L}(\mathscr{T}) = \sum_t\sum_{h \not\in \mathscr{T}} log(Poiss_{n_j^{(t)}}(n^{(t)}\epsilon^{x(h)}))
\end{equation}

Here $t$ indexes the time point, $n^{(t)}$ and $\epsilon=0.003$ are the total depth and the error rate, respectively. The values $x(h)$ represents the number of mutants in haplotype $h$. The tree with maximum likelihood is selected.

{\bf Step 2} \emph{Supertree Construction} We next build supertrees of the evolution from the subtrees. As we saw in the example in Figure \ref{PedagogicSample}F, this involves ranking the subtree branches by prevalence, and adding mutations sequentially as in the example in Figure \ref{PedagogicSample}F, checking pairwise ancestry relationships between mutations (from the subtrees), along with the capacity of prevalence available at each node (by checking equation (1) for every time point).

Note that this algorithm may produce no trees. This implies there are no supertrees consistent with the data, and recombination networks may be more suitable.


\subsection*{The Recombination Algorithm}

We would like to use data such as Figure \ref{NetworkFigure}B to reconstruct the evolutionary structure. The splits method \cite{Huson} is used to construct phylogenetic networks such as Figure \ref{NetworkFigure}G. There are many recombination networks that correspond to any given phylogenetic network. A standard method to identify recombination networks is to look for an optimal path of trees across the recombination sites \cite{Song}. These methods generally have the full mutation profile of a set of species of interest to compare. Our problem is exacerbated by missing data and the full haplotypes of distinct species (clones in our case) are not available. However, we have prevalence information which can help identify structures consistent with the data.

We construct recombination networks in five steps; haplotype classification, super-network construction, super-network filtering, prevalence maximum likelihood estimation, and prevalence range estimation. We describe these steps in detail.

{\bf Step 1} \emph{Haplotype Classification}. In order to distinguish the real and artifact haplotypes in any table such as Figure \ref{MiSeqHiSeq2}C we do the following. For any count $n_h^{(t)}$ associated with haplotype $h$ and time point $t$, we calculate the probability it arises as an error using the Poisson distribution. This gives a term of the form $Poiss_{n_h^{(t)}}(n^{(t)}\epsilon^{x(h)})$, where $n^{(t)}$ is the total read depth from that time point, $x(h)$ is the number of mutations distinct from the wild type, and $\epsilon$ is a user selected error rate per base per read. We then take the combined log-likelihood $\mathcal{L}$ across all time points. All log-likelihoods below a threshold $\mathcal{L}_{0}$ are classified as artifact. The values $\epsilon = 0.005$ and $\mathcal{L}_{0}=-360$ were used in implementation.

{\bf Step 2} \emph{Super-Network Construction}. This is a brute force approach where we construct all possible recombination networks using $r=0,1,2,...$ reticulated nodes in turn. Any networks that do not contain the real haplotypes of the individual haplotype tables of Step 1 are rejected. The value of $r$ selected is the smallest value with any valid networks after Steps 3 and 4 are implemented.

{\bf Step 3} \emph{Filtering}. We need to utilize the prevalence to identify and remove invalid networks. Each leaf $c$ of the recombination graph represents a single clone of the mixed population. We let $\rho_c$ denote the prevalences of that clone. We then have the conditions:

\begin{equation}
\sum\limits_c \rho_c=1,\rho_c \ge 0
\label{Simplex}
\end{equation}

Now we have the estimated prevalence $\lambda_m$ of each mutation $m=1,2,...,M$ from the proportional sequencing depth at the mutations position. If we let $\mathcal{C}_m$ denote the set of clones from the super-network that contain mutation $m$, we have conditions of the form:

\begin{equation}
\sum\limits_{\{c \in \mathcal{C}_m\}}\rho_c =\lambda_m
\label{Filter}
\end{equation}

We solve the linear programming problem defined by Equations \ref{Simplex} and \ref{Filter} with the simplex method. If no solution exists on any day $t$ the network is rejected. If a solution is found, it is the input to the (more precise) calculation in Step 4. This step generally reduces the number of networks to manageable levels.

{\bf Step 4} \emph{Prevalence Point Estimation} In reality $\lambda_m$ is an estimate and we have more information than just the depth of mutations. For each cluster of mutations we have the count $n_h^{(t)}$ for each real genotype $h$ (artifacts are ignored) and time points $t$ in the corresponding table. Conditioning on the total count $n^{(t)}$ of real genotypes results in a binomial log-likelihood of the following form:

\begin{center}
$\mathcal{L}=\sum\limits_hn_h^{(t)}log(n^{(t)}\sum\limits_{c \in C_h}\rho_c^{(t)})$
\end{center}

Here the sum is over the set $C_h$ of clones that contain haplotype $h$. We then sum this over all tables and time points and maximize for estimates of the clone prevalences $\rho_c^{(t)}$. We use gradient descent to maximize, projecting each step onto the simplex in Equation \ref{Simplex}. Projecting onto the simplex is relatively straightforward, the updated prevalence vector {\bf $\rho$} just becomes $\frac{{\bf \rho}}{||{\bf \rho}||_1}$, where negative components are set to zero.

{\bf Step 5} \emph{Range Estimation}. Step 4 does not always result in a unique estimate, because there may be ranges of values $\rho_c^{(t)}$ on the simplex of Equation \ref{Simplex} that yield identical terms $\sum\limits_{c \in s}\rho_c^{(t)}$. Then if  $\hat\rho_c^{(t)}$ are the estimates from the gradient descent, we use the simplex method to maximize $\pm \rho_c^{(t)}$ subject to Equation \ref{Simplex} and conditions of the form:

\begin{center}
$\sum\limits_{c \in s}\rho_c^{(t)}=\sum\limits_{c \in s}\hat\rho_c^{(t)}$
\end{center}

Valid clonesets with the maximum likelihood are then selected. This can be applied to any putative network to either conclude that the network is not feasible, or produce a range of possible prevalences associated with the network.


\subsection*{Template Switching}

We model template switching during rtPCR as follows. Suppose we have two mutations of interest and four possible genotypes, labeled $C_{00}$, $C_{01}$, $C_{10}$ and $C_{11}$. We have corresponding read depth counts $n_{00}$, $n_{01}$, $n_{10}$ and $n_{11}$.  Now, if tree like evolution exists, one of $C_{01}$, $C_{10}$ or $C_{11}$ is an artifact arising from template switching during rtPCR (the wild type C00 is assumed to always occur). We demonstrate the case where $C_{01}$ is an artifact (model 2 in Figure \ref{TemplateSwitching}Aii). The derivation for the other two models is similar. Then we assume that the real clones $C_{00}$, $C_{10}$ and $C_{11}$ have prevalences of $a$, $b$ and $c$, respectively, so that $a+b+c=1$.

We model rtPCR as a time continuous three state process, where template switching occurs at a rate $\lambda$, jumping to any of the three templates $C_{00}$, $C_{10}$ or $C_{11}$ with probabilities $a$, $b$ and $c$, respectively. We also refer to the states as $a$, $b$ and $c$. We let $p_a(t)$, $p_b(t)$ and $p_c(t)$ be the probabilities of occupying a copy of the corresponding templates at position $t$. Then conditioning $p_a(t)$ over a time interval $(t,t+dt)$ results in the following expression (see \cite{Bunday} for typical derivations):

\begin{center}
$p_a(t+dt)=p_a(t)(1-\lambda dt) +p_a(t)a\lambda dt+p_b(t)a\lambda dt+p_c(t)a\lambda dt$
\end{center}

This gives us the following differential equation and solution:

\begin{center}
$\frac{dp_a}{dt}=\lambda(a- p_a) \iff p_a(t)=a+(p_a(0)-a)e^{-\lambda t}$
\end{center}

We rescale time so that $t=1$ represents one rtPCR cycle. We then have the following transition matrix between states:

\begin{center}
T=$\left( \begin{array}{ccc}
a+(1-a)e^{-\lambda} & b-be^{-\lambda} & c-ce^{-\lambda} \\
a-ae^{-\lambda} & b+(1-b)e^{-\lambda} & c-ce^{-\lambda} \\
a-ae^{-\lambda} & b-be^{-\lambda} & c+(1-c)e^{-\lambda} \end{array} \right)$
\end{center}

Probabilities for all types $C_{00}$, $C_{01}$, $C_{10}$ and $C_{11}$ can now be defined, which we demonstrate for $C_{10}$. Derivations for the other terms can be obtained in a similar manner. From Figure \ref{TemplateSwitching}Aii we see that to obtain a read of the form $C_{10}$, we can start in either state $b$ or $c$ and end in either state $a$ or $b$. This gives us four terms to add:

\begin{center}
$Pr(C_{10}) = bT_{ba}+bT_{bb}+cT_{ca}+cT_{cb}=b(a-ae^{-\lambda}+b+(1-b)e^{-\lambda})+c(a-ae^{-\lambda} + b-be^{-\lambda})=b+acn$
\end{center}

Here $n=1-e^{-\lambda}$ is the probability a template switch occurs. The formulas  in Figure \ref{TemplateSwitching}A are obtained similarly.

The counts $n_{00}$, $n_{01}$, $n_{10}$ and $n_{11}$ then follow a multinomial distribution, from which log-likelihoods can be derived. A chi-squared goodness of fit can then be obtained. We note that in many cases, solutions for the four terms $Pr(C_{00})$, $Pr(C_{01})$, $Pr(C_{10})$ and $Pr(C_{11})$ in terms of $a$, $b$, $c$ and $n$ can be obtained, resulting in a perfect fit. When this is not possible, one or more of the three models can be rejected if the fit is sufficiently bad.

Note that none of these three models necessarily explain the data. In the last column of Figure \ref{TemplateSwitching}D, for example, we have four artificial counts $50$, $1000$, $1000$ and $1000$ corresponding to genotypes $C_{00}$, $C_{01}$, $C_{10}$ and $C_{11}$. All three models are a bad fit suggesting recombination is present. However, this relies on small counts for $C_{00}$, which were not observed in the real data that was examined.

Note that template switching has no effect on the prevalence of individual mutations. For example, considering Figure \ref{MiSeqHiSeq}Ciii, if we add $Pr(C_{01})$ and $Pr(C_{11})$, we get $b+c$, which is precisely the prevalence of mutation $M2$.


\subsection*{Validation and Simulation}
\label{Val}

The validation of the method is based upon simulated data. This will give some idea of the reconstruction capabilities of the methods and allow benchmarking with other existing approaches. In particular, we compared our tree construction algorithm to the benchmark software Shorah using the same simulation approaches as Zagordi et al \cite{zagordi2010deep,zagordi2011shorah} and Astrovskaya et al. \cite{astrovskaya2011inferring}.

To measure the performance of the mixed population estimation, we computed the \emph{Precision}, the \emph{Recall}, and the \emph{Accuracy} of prevalence estimation for the methods of interest. The recall (or sensitivity) gives the ratio $\frac{TP}{TP+FN}$ of correctly reconstructed haplotypes to the total number of true haplotypes, where we have true positives ($TP$), false negatives ($FN$) and false positives ($FP$). The precision gives their ratio to the total number of generated haplotypes, $\frac{TP}{TP+FP}$.

The accuracy measures the ability of the method to recover the true mixture of haplotypes, and was defined as measuring the mean absolute error of the prevalence estimate. Where a range estimate is obtained for the prevalence, we calculate the shortest distance from the true value to the range.

\begin{figure}[t!]
\centering
\includegraphics[width=14cm]{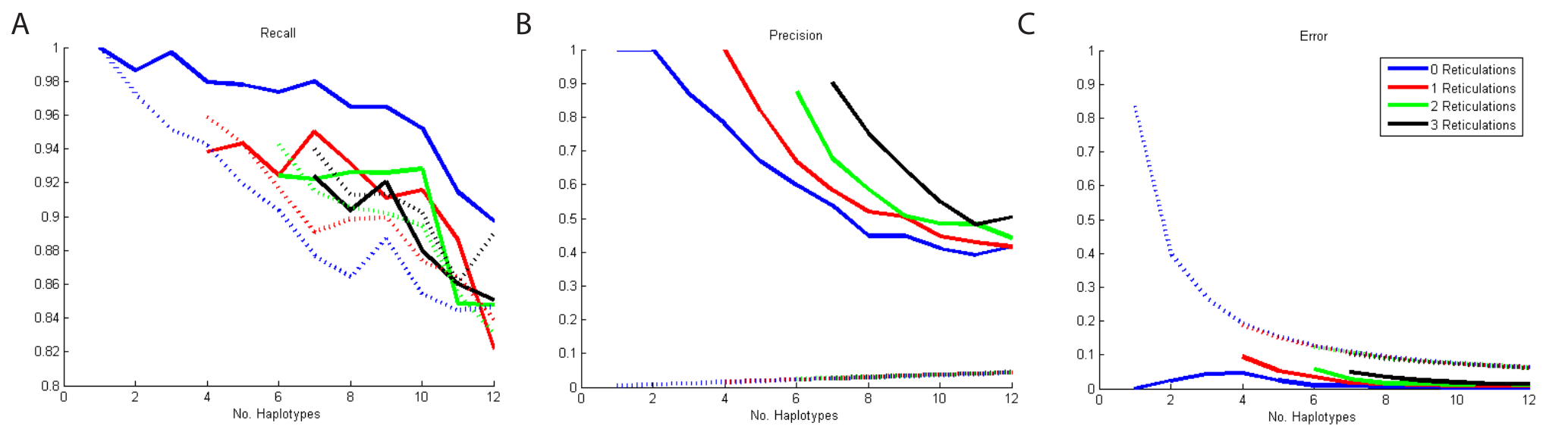}
\caption{Validation profiles for a range of haplotype counts, including; (A) Recall, (B) Precision, and (C) Error. In all cases the solid line denotes the algorithm preformance, the dashed line indicates Shorah performance.}
\label{Validation}
\end{figure}

Comparison with Shorah was done on simulated deep sequencing data from a 1.5 kb-long  region of HIV-1. Simulated reads have been generated by MetaSim\cite{richter2011metasim}, a meta-genomic simulator which generates collections of reads reproducing the error model of some given technologies such as Sanger and 454 Roche. It takes as input a set of genome sequences and an abundancy profile and generates a collection of reads sampling the inputted genomic population.

For up to $12$ haplotypes and $3$ reticulations we performed 100 runs as follows. We randomly constructed a network by attaching each new branch to a random selected node. Reticulations were also randomized. The prevalences of the resulting clones (at the leaves) were randomly selected from a Dirichlet distribution. This is repeated for 10 time points of data. We used MetaSim to generate a collection of 5,000 reads having an average length of 500bp and replicating the error process of Roche 454 sequencing. The methods were then applied to the resulting data.

Shorah output can display mismatches or gaps in the outputted genomes, with increasing frequency at the segment edges. We applied a modification on Shorah output by trimming the edge and we then corrected one or two mismatches or gaps on all the genomes before addressing the comparison. Figures \ref{Validation}A-C provide the comparison for recall, precision and error indicators. We found slight improvements for recall, especially for tree like evolution. The precision and error also had improved results. We acknowledge that the simulations were based upon evolutionary structures that the models are designed to fit so such improvement might be expected. Furthermore, Shorah likely have better performance on low prevalence clones. However, these simulations demonstrate that reasonable results can be obtained from the techniques we have introduced.


\footnotesize{

}
\end{document}